\documentclass[3p,times,procedia]{elsarticle}
\usepackage{ecrc}

\volume{00}

\firstpage{1}

\journalname{23rd  EURO Working Group on Transportation Meeting, EWGT 2020, 16-18 September 2020, Paphos, Cyprus}

\runauth{A. Georgantas et al.}

\jid{trpro}

\jnltitlelogo{Transportation Research}

\usepackage{amssymb}
 \usepackage{amsthm}

\biboptions{authoryear}

\usepackage[figuresright]{rotating}

\usepackage[bookmarks=false]{hyperref}
    \hypersetup{colorlinks,
      linkcolor=blue,
      citecolor=blue,
      urlcolor=blue}

\newcommand{\vecr}{\mbox{\boldmath $\rho$}}
\newcommand{\vecq}{\mbox{\boldmath $q$}}
\newcommand{\vecrr}{\mbox{\boldmath $r$}}
\newcommand{\vecx}{\mbox{\boldmath $\xi$}}

\newcommand{\vecpsi}{\mbox{\boldmath $\psi$}}

\newcommand{\vecy}{\mbox{\boldmath $y$}}

\begin{document}
\begin{frontmatter}




\title{Route Reservation Architecture in Tandem Transportation Networks with the Asymmetric Cell Transmission Model}


\author[a]{A. Georgantas}
\author[a]{C. Menelaou}
\author[a]{C. Panayiotou}

\address[a]{University of Cyprus, Department of Electrical and Computer Engineering, 1 Panepistimiou Avenue, Nicosia, 1678, Cyprus}

\begin{abstract}

Traffic congestion in big cities has been proven to be a difficult problem with suboptimal effects in terms of driver delay and frustration, cost and impact on the environment.~In principle, many transportation networks lack a unified framework, which will coordinate the traffic in such a manner, in order to suppress congestion and at the same time improve the travel time of the users situated in it.~The rapid advancements in information, communication, and computation technologies have given rise to more elaborate modeling frameworks, aiming to act as the coordination unit necessary to counter the issue of congestion in real-time conditions.~Such actions might have an adverse effect on the efficiency of the network, prospectively leading to greater waiting time intervals for each individual driver.~We propose a macroscopic model equipped with an underlying reservation feature, known as Route Reservation Architecture~(RRA).~Vehicles enter the network~(mainstream-wise or from the on-ramps) as long as this inflow does not incur a density that exceeds the network's critical density.~Those vehicles prospectively exceeding the critical density are stored as queues at the origin of the motorway stretch and within the on-ramps~\cite{Kotsialos2004}.~Once there is sufficient space, for those vehicles to be accommodated by the respective cell of the network, they are discharged from their queueing instance at their respective origin, moving towards their assigned path freely.~In previous works, this architecture has been only applied in microscopic simulations in the context of urban networks without the influence of source terms~(on-ramps)~\cite{Menelaou2017}.~When the critical density of the stretch is reached, the reservations are activated, instigating a waiting interval to the vehicles stored at queues, allowing vehicles only to enter the network at a later time instant, such that the critical density is not crossed.~In this vein, we avoid the congested region and operate only at free-flow conditions.~Our target is to investigate the effectiveness of this architecture, i.e. to minimize the Total Travel Time (TTT) of the vehicles present in the network, along with the vehicles situated in the queues, through a macroscopic simulation framework.






\vspace{-0.3cm}



\end{abstract}

\begin{keyword}
critical density \sep route reservations \sep travel time spent \sep cell transmission model
\vspace{-0.3cm}
\end{keyword}

\cortext[cor1]{Corresponding authors: A. Georgantas, C. Menelaou, C. Panayiotou}

\end{frontmatter}
\vspace{-0.9cm}
\email{\{georgantas.antonios, menelaou.charalambos, christosp\}@ucy.ac.cy}

\section{System Model and Problem Statement}
\label{main}


We consider a tandem network that consists of $\mathcal{P} = \{1,2,\ldots,m,\ldots, M\}$ cells denoted as $1$ through $M$, starting from the upstream-most section.~Each cell may consist of an on-ramp and an off-ramp and operates according to the \textit{Fundamental Diagram}~(FD)~\cite{Knoop2017IntroductionTT}.~We use a form of interaction among these discretized cells inspired by the \textit{Asymmetric Cell Transmission Model~(ACTM)}~\cite{Gomes2006}, which poses a modified version of Daganzo's original Cell Transmission Model (CTM)~\cite{Daganzo1994}.



%
\vspace{-0.2cm}

\subsection{Traffic Dynamics}

\vspace{-0.4cm}




\noindent We consider the following discrete-time equations that dictate the evolution of the dynamics of the system, as follows:

\begin{equation}\label{fixx}
\hspace*{-1.0cm}\vecr_m(k+1) = \vecr_m(k) + \frac{T}{L_m}\Bigg(\vecq_{m-1}(\vecr_{m-1}(k),\vecr_{m}(k),\vecrr_{m+1}(k))~-~ \frac{\vecq_{m}(\vecr_{m}(k),\vecr_{m+1}(k),\vecrr_{m+1}(k))}{1-b_m} + \vecrr_m(k)\Bigg),  \forall m\in \mathcal{P} - \{1,M\}\\[6pt]
\end{equation}

\noindent where $T$ is the simulation time step, $\vecr_{m}(k)$ stands for the density of $m$-th cell at time $k$, $\vecrr_m(k)$ denotes the on-ramp flow, $\vecq_{m}(\vecr_{m}(k),\vecr_{m+1}(k),\vecrr_{m+1}(k))$ the flow exiting from $m$-th heading towards cell $m+1$.~Additionally,~$L_m$ is the cell's length and $b_m$ is the percentage of flow exiting from the off-ramp attached to cell $m$.



\vspace{-0.5cm}

\section{Reservation Dynamics}

\vspace{-0.2cm}


We denote $\vecpsi_m(k)$~(veh/h) as the route reservation vector of vehicles assigned to enter cell $m+1$ originating from $m$.~Similarly, $\vecy_{m}(k)$~(veh/h) is the reservation vector acting on each present on-ramp controlling the traffic volume that will be inserted at the merging region of cell $m$.

Let $n_m(k)$ denote the accumulated number of traffic flow reservations~(veh/h) corresponding to each cell $m$ at time step $k$.~Each cell $m$ is characterized as admissible as long as the flow of vehicles entering at time step $k$~(either from the origin of the network or the attached on-ramp instances) can traverse cell $m$ without exceeding its underlying critical density, $\rho_{c,m}$ during any time instant for which the explicit traffic flow of vehicles would be travelling at the cell during the interval $[k,k+\Delta k]$, $\Delta k$ being the traversal time of cell $m$.~Thus,~quantity $\displaystyle \frac{T}{L_m}(1-b_m)n_m(k)$ denotes the instantaneous density associated with cell $m$ at time step $k$.~The admissibility state for cell $m$ at time step $k$ is denoted as $\vecx_m(k)$,~where $\vecx_m(k)$ is equal to $1$, if cell $m$ is indeed admissible or $0$, if it is not permissible for vehicles to enter inside it.~More explicitly,

\begin{equation}\label{eq:admissible}
\displaystyle \vecx_m(k)= \left\{
\begin{array}{lll}
1, & \textrm{if }~ \displaystyle \frac{T}{L_m}(1-b_m)n_m(k+\zeta) \leq \rho_{c,m}, ~\forall \zeta\in [k,k+\Delta k]~\\[6pt]
0, &\textrm{otherwise}
\end{array} 
\right. 
\end{equation}

Vehicles are allowed to traverse road segments during those instants, in which $\vecx_m(k)=1$, since we can enable the necessary reservations to ensure congestion-free routing.~In the case, where vehicles enter from the on-ramps, then this reduced value reads as $\vecy_m(k) = h_m(k)\vecrr_m(k)$.~Similarly, in the case where vehicles move from cell to cell with respect to the mainstream front, the delay is reflected in $\vecpsi_m(k)=(1-b_{m-1})\beta_m(k)\vecq_{m-1}(\cdot)$, where $\beta_m$~($\in[0,1]$) is the percentage of the nominal flow that can actually leave from $m$ and relocates to $m+1$.~In the case of route reservations, the traffic dynamics shown in (1) depend inherently from the admissibility state $\vecx_m(k)$ and the reservation vectors $\vecy_m(k),\vecpsi_m(k)$, so Eq. (1) is adjusted accordingly.

\vspace{-0.4cm}

\section*{Acknowledgements}
This work has been supported by the European Union's Horizon 2020 research and innovation programme under grant agreement No. 739551 (KIOS CoE) and from the Government of the Republic of Cyprus through the Directorate General for European Programmes, Coordination and Development.

\bibliography{PaperGeorgantas}
\bibliographystyle{elsarticle-harv}

\clearpage

\normalMode

%
%
%
%

\end{document}